\documentclass[useAMS,usenatbib]{mn2e}
\usepackage{txfonts}
\usepackage{natbib}
\usepackage[all]{xy}
\usepackage{graphicx}
\usepackage{multirow}

\usepackage{epstopdf}
\usepackage[latin1]{inputenc}
\usepackage{scrpage2}
\usepackage{rotating}
\usepackage{subfigure}
\usepackage{wrapfig}
\usepackage{subfig}
\usepackage{subfloat}
\usepackage[english,ngerman]{babel}

\def\del#1{{}}

\newcommand{\ltsima}{$\; \buildrel < \over \sim \;$}
\newcommand{\lsim}{\lower.5ex\hbox{\ltsima}}
\newcommand{\gtsima}{$\; \buildrel > \over \sim \;$}
\newcommand{\gsim}{\lower.5ex\hbox{\gtsima}}
\newcommand{\bra}{\langle}
\newcommand{\ket}{\rangle}

\newcommand{\f}{\frac}

\newcommand{\bc}{\begin{center}}
\newcommand{\ec}{\end{center}}

\title[relative velocities with PINOCCHIO]
{A study of relative velocity statistics in Lagrangian perturbation theory with PINOCCHIO}
\author[L. Heisenberg, B.M.  Sch\"afer and M. Bartelmann]
{Lavinia Heisenberg\thanks{e-mail: lavinia.heisenberg@unige.ch}$^{1,2}$, Bj{\"o}rn Malte Sch\"afer$^{3}$ and Matthias Bartelmann$^2$\\
$^1$ Universit{\'e} de Gen{\`e}ve, D{\'e}partement de Physique Th{\'e}orique, 24, quai E. Ansermet, 1211 Gen{\`e}ve, Switzerland\\
$^2$ Institut f{\"u}r theoretische Astrophysik, Zentrum f{\"u}r Astronomie, Universit{\"a}t Heidelberg, Albert-Ueberle-Stra{\ss}e 2, 69120 Heidelberg, Germany\\
$^3$ Astronomisches Recheninstitut, Zentrum f{\"u}r Astronomie, Universit{\"a}t Heidelberg, M{\"o}nchhofstra{\ss}e 12, 69120 Heidelberg, Germany}

\begin{document}
\pagerange{\pageref{firstpage}-\pageref{lastpage}}
\pubyear{2008}
\maketitle
\label{firstpage}

\begin{abstract}
Subject of this paper is a detailed analysis of the PINOCCHIO algorithm for studying the relative velocity statistics of merging haloes in Lagrangian perturbation theory. Given a cosmological background model, a power spectrum of fluctuations as well as a Gaussian linear density contrast field $\delta_{\rm l}$ is generated on a cubic grid, which is then smoothed repeatedly with Gaussian filters. For each Lagrangian particle at position $\bmath{q}$ and each smoothing radius $R$, the collapse time, the velocities and ellipsoidal truncation are computed using Lagrangian Perturbation Theory. The collapsed medium is then fragmented into isolated objects by an algorithm designed to mimic the accretion and merger events of hierarchical collapse. Directly after the fragmentation process the mass function, merger histories of haloes and the statistics of the relative velocities at merging are evaluated. We reimplemented the algorithm in C++, recovered the mass function and optimised the construction of halo merging histories. Comparing our results with the output of the Millennium simulation suggests that PINOCCHIO is well suited for studying relative velocities of merging haloes and is able to reproduce the pairwise velocity distribution.
\end{abstract}


\section{Introduction}
To account for structure formation one needs to develop techniques for studying the nonlinear evolution of perturbations. In the strongly nonlinear regime where the perturbation amplitudes exceed unity ($|\delta|\gg 1$), the linear approximation breaks down and has to be replaced by other approaches. To accentuate the need for this, one has to look at the interesting structures in the Universe, like galaxies or clusters of galaxies because they are highly non-linear. Since every theory of structure formation must be capable of describing the formation and evolution of non-linear objects, the major developments occurred in perturbation theory and numerical simulations. It has been understood that this aim simplifies when formulated in terms of Lagrangian coordinates rather than the standard Eulerian ones, since the latter one relies on physical densities being small \citep{1992ApJ...394L...5B,1992MNRAS.254..729B}. So one assumes that the dynamics of gravitational clustering is described  more suitably in terms of the displacement field $\bmath{D}$, which is in the Lagrangian approach the only underlying fundamental field. The decisive difference to the Eulerian approach is that it is not based on the smallness of the density of the inhomogeneities and that one searches for solutions of perturbed trajectories about the linear initial displacement $\bmath{D}^{(1)}$ \citep{2002PhR...367....1B,1997GReGr..29..733E,1995PhR...262....1S}. The fundamental point is that a small perturbation of the Lagrangian particle paths carries a large amount of non-linear information about the corresponding Eulerian evolved observables, since the Lagrangian picture is intrinsically non-linear in the density field. Lagrangian perturbation theory reaches its limit of applicability when trajectories of particles cross and the mapping from the initial conditions to the evolved density field ceases to be unique. This defines the so-called orbit-crossing which is numerically achieved by means of the ellipsoidal collapse approximation to the full Lagrangian perturbative expansion. These collapsed objects can then be grouped into disjoint haloes and their relative velocities can be measured.

There is a number of applications of large simulated volumes with velocity information: For instance, simulations of redshift-space distortions in the galaxy correlation function, investigations of large-scale bulk flows \citep{2002MNRAS.336.1234Z} which offer a possibility of testing modified gravity models \citep{2009arXiv0908.2903A}, and testing the occurence of high-velocity merging events, such as the bullet-cluster, which is the topic of recent controversies \citep{2006MNRAS.370L..38H}. Many cluster observables such as the strong-lensing cross-section \citep{2006A&A...447..419F,2004MNRAS.349..476T} and the radio and X-ray luminosity \citep{2002ApJ...577..579R} are boosted during the merging process, and for treating these systems, statistical information about the merging configuration is needed.

The cosmological model used is a spatially flat $\Lambda$CDM cosmology with Gaussian adiabatic initial perturbations in the cold dark matter (CDM) density field with the following cosmological parameter set $(\Omega_{m0}, \Omega_{\Lambda}, h, \sigma_8)=(0.25, 0.75, 0.73, 0.9)$, identical to that used in the Millennium simulation, to which we will compare our results. Our computational domain is a cubic box of size $256~\mathrm{Mpc}/h$ with periodic boundary conditions filled with $N^3=128^3$ particles.

\section{cosmology}\label{sect_cosmology}\cite{}

In this first introductory section we summarize all the known formulas we use to compute the power spectrum and the Gaussian linear density contrast field.
The linear CDM power spectrum $P(k)$ describes the fluctuation amplitude of the Gaussian initial density field $\delta$, $\bra\delta(\bmath{k})\delta(\bmath{k}^\prime)\ket=(2\pi)^3\delta_D(\bmath{k}+\bmath{k}^\prime)P(k)$, and is given by the ansatz
\begin{equation}
P(k)\propto k^{n_s}T^2(k),
\end{equation}
with the transfer function $T(k)$ which is approximated by \citep{1986ApJ...304...15B},
\begin{equation}
T(q) = \frac{\ln(1+aq)}{aq}\left(1+bq+(cq)^2+(dq)^3+(eq)^4\right)^{-\frac{1}{4}}.
\label{eqn_cdm_transfer}
\end{equation}
In the transfer function, the wave number $k=q\Gamma$ enters rescaled by the shape parameter $\Gamma$ \citep{1995ApJS..100..281S},
\begin{equation}
\Gamma=\Omega_m h\exp\left(-\Omega_b\left(1+\frac{\sqrt{2h}}{\Omega_m}\right)\right).
\end{equation}

For further treatment one introduces the smoothed density field $\delta_R$, which is averaged on the scale $R$,
\begin{equation}
\tilde{\delta}_R(\bmath{k})=\tilde{\delta}(\bmath{k})\tilde{W}_R(\bmath{k}),
\end{equation}
and for the smoothed power spectrum
\begin{equation}
P_R(k)=|\tilde{W}_R(k)|^2P(k).
\end{equation}
with the window function $W_R(r)$, satisfying $\int d^3x\; W_R(|\bmath{x}|)=1$. 
There exist several choices for the window function $W_R$, a very common one is the top-hat filter, which is given by 
\begin{equation}
\tilde{W}_R(k)=3\;\frac{\sin{(kR)}-kR\cos{(kR)}}{(kR^3)}.
\end{equation}
Using the definitions introduced so far, it is now straightforward to calculate the variance $\sigma^2(R)=\langle\delta_r^2(\bmath{x})\rangle$ of the smoothed density field as
\begin{equation}\label{eq:smooth_variance}
\sigma^2(R)=\int\frac{d^3k}{(2\pi)^3}\;P_R(k)=\int\frac{d^3k}{(2\pi)^3}\;|\tilde{W}_R(k)|^2P(k).
\end{equation}
which we use to fix the normalization to $\sigma^2(R=8\:\mathrm{Mpc}/h)\equiv\sigma^2_8$.

\section{Lagrangian Perturbation theory}\label{sect_lagrangian}
In this section we summarize the idea and results of the Lagrangian perturbation theory, which we use to study the dynamical equations of the considered non-linear system. The formation of CDM haloes involves highly non-linear dynamical processes which can only be followed in numerical simulations or by applying perturbative methods. As mentioned in the introduction, one important approach is Lagrangian Perturbation Theory. In order to describe the non-linear dynamical evolution of the particle trajectories up to third order Lagrangian coordinates the following three steps have to be followed:
\begin{enumerate}
 \item Description of the mapping from the Eulerian to the Lagrangian coordinates
 \item Transformation of the Eulerian fields to Lagrangian coordinates
 \item Application of perturbation theory to the Lagrangian equations expressed by the displacement up to desired order
\end{enumerate}
The starting point is a pressureless, non-vortical, self-gravitating fluid with Newtonian gravity embedded in an expanding Friedmann-Lema{\^i}tre Robertson-Walker universe \citep{1992ApJ...394L...5B}.
In the Lagrangian framework of fluid dynamics the relation between the Eulerian position $\bmath{x}$ of a mass particle and the initial Lagrangian position $\bmath{q}$ is given by the displacement field \citep{1970A&A.....5...84Z}:
\begin{equation}\label{eq:displacement}
 {\bmath{x}}({\bmath{q}},t) = {\bmath{q}} + {\bmath{D}}({\bmath{q}},t),
\end{equation}
In the Lagrangian space the trajectories of the mass elements are fully described by the dynamical mappings $\bmath{x}(\bmath{q},t)$, starting from the initial positions $\bmath{q}$.\\ 
There is a one-to-one correspondence between the Langrangian coordinate $\bmath{q}$ and the Eulerian coordinate $\bmath{x}$ for a cold non-collisional fluid, at least until the stage of caustic formation. Expressed mathematically, this corresponds to the statement that the functional determinant $J$ of the Jacobian of the mapping relation $\bmath{q} \rightarrow \bmath{x}(\bmath{q},t)$ is non-singular,
\begin{equation}
J(\bmath{q}, t) \equiv {\rm det}\left(\f{\partial\bmath{x} }{\partial\bmath{q}}  \right) \neq 0\;,
\end{equation} 
which means that the mapping $\bmath{x}(\bmath{q},t)$ is invertible to $\bmath{q}(\bmath{x},t)$. It is evident that many particles coming from very different original positions will tend to arrive at the same Eulerian position during the highly non-linear evolution. As a consequence of that infinite-density regions (caustics) will form in Eulerian space. Hence the mapping from Lagrangian to Eulerian space becomes singular and the density infinite as $\rho\propto J^{-1}$. Since the displacement $\bmath{D}$ fully characterizes the map between the Eulerian and the Lagrangian coordinates, the motion of the fluid elements are completely described in terms of it. One can now express the peculiar velocity, acceleration and density contrast by inserting the displacement field $\bmath{D}$ into the Euler and the continuity equations. Similarly the Eulerian irrotationality condition and the Poisson equation can be written in terms of the displacement \citep{1995MNRAS.276..115C} resulting in the dynamical equations for the displacements $\bmath{D}$. The fundamental question now is how to solve the dynamical equations for the displacements $\bmath{D}$. The irrotationality condition and the Lagrangian Poisson equation are exact equations in the Lagrangian description. It is undoubtedly very difficult to solve them in a rigorous way. A possible alternative is to seek for approximate solutions. The standard technique is to expand the trajectory $\bmath{D}$ in a perturbative series, the leading term being the linear displacement which corresponds to the Zel'dovich approximation \citep{1992ApJ...394L...5B,1970A&A.....5...84Z}. Approximating the Lagrangian Poisson equation implies that the gravitational interaction among the particles of the fluid is described by the first few terms of a Taylor expansion.
\begin{equation}\label{eq:perturDispl}
\bmath{D}(\bmath{q},\tau) =
g_1(\tau)\,\bmath{D}^{(1)}(\bmath{q})
+g_2(\tau)\,\bmath{D}^{(2)}(\bmath{q}) + \cdots\;
\end{equation}
$\bmath{D}^{(n)}(\bmath{q})$ being the $n$th-order approximation. 
The dynamics of the evolution constrains in general both the temporal dependence as described by the functions $g_n$, and the spatial displacements $\bmath{D}^{(n)}(\bmath{q})$.
The displacement fields are computed considering the potential of a homogeneous ellipsoid in its principal axis frame $\psi(\bmath{q})=\frac{1}{2}(\lambda_1q_1^2+\lambda_2q_2^2+\lambda_3q_3^2)$ solving the Poisson and irrotationality condition equtions in Lagrangian space, where the $\lambda_i$ are the eigenvalues of the first-order deformation tensor $\psi_{,ab}(\bmath{q})$. The solutions can be found in \citep{1995MNRAS.276..115C} and 
\citep{1997MNRAS.287..753M,1997MNRAS.290..439M} to which we refer for all details. We use the results from \citep{1997MNRAS.287..753M} for the calculation of collapse times $J(\bmath{q},g_c)=0$
\begin{equation}\label{eq:collapseTimes}
1+\lambda_i g_c - \frac{3}{14} \lambda_i (\delta_l-\lambda_i)
g_c^2 + \left( \frac{I_3}{126} + \frac{5}{84}\lambda_i
\delta_l(\delta_l-\lambda_i)\right)g_c^3= 0.
\end{equation}
Just in order to give some examples, the first order collapse time is given by $g_c^{(1)}=-\frac{1}{\lambda_3}$ whereas the second order solution results to be
$g_c^{(2)} = \frac{7\lambda_3 + \sqrt{7 \lambda_3(\lambda_3+6\delta_l)}}
{3\lambda_3(-\lambda_3+\delta_l)}\;$ \citep[and analogously for the third order, see ][]{1997MNRAS.287..753M}.

 
\subsection{PINOCCHIO}
A compromise between simulations and analytical techniques is a perturbative approach describing the growth of haloes in a given numerical realization of a linear density field, such as the truncated Zel'dovich approximation and the PINOCCHIO algorithm \citep{2002MNRAS.333..623T}. PINOCCHIO (acronym for {\em PINpointing Orbit-Crossing Collapsed HIerarchical Objects}) is an algorithm for studying the formation and evolution of dark matter haloes in a given initial linear density field, which we outline briefly in this section. It was first developed by \citet{2002MNRAS.331..587M}. Local parameterisations to the dynamics are used to give precise predictions of the hierarchical formation of dark matter haloes when the correlations in the initial density field are properly taken into account.

This modus operandi enables the automatic generation of a large ensemble of accurate halo merging histories and additionally delivers their spatial distribution. Likewise, the approach can be efficiently applied for generating the input for galaxy formation models since the properties of the halo population are of fundamental importance for understanding galaxy formation and evolution.

PINOCCHIO consists of two steps which determine the hierarchical formation of haloes through accretion and merging: 
\begin{itemize}
 \item The first step handles the definition of the collapse time. Hereby, orbit-crossing will be identified as the instant when a mass element undergoes collapse, without the need to introduce a free parameter. Orbit-crossing is numerically calculated by means of the ellipsoidal collapse approximation to the full Lagrangian perturbative expansion, as discussed in the previous section \citep{1995ApJ...447...23M}. 
 \item The second step groups the collapsed particles into disjoint haloes, applying an algorithm similar to that used to identify haloes in $n$-body simulations.
\end{itemize}
As explained in the previous subsection, the density diverges as the Jacobian determinant vanishes, corresponding to the formation of a caustic, which means that particle trajectories cross and the transformation ${\bmath{x}}\rightarrow{\bmath{q}}$ becomes multi-valued. Since the density becomes very high at orbit crossing, this event will be identified as the collapse time. In this manner, collapse is very easy to compute using Lagrangian Perturbation Theory which remains valid up to that particular instant but breaks down afterwards.

\begin{figure}
\begin{center}
\begin{tabular}{c}
\includegraphics[width=0.35\textwidth]{./case1.eps} \\
\includegraphics[width=0.35\textwidth]{./case2.eps} \\
\includegraphics[width=0.35\textwidth]{./case3.eps} \\
\includegraphics[width=0.35\textwidth]{./case4.eps}
\end{tabular}
\end{center}
\caption{Cases of the fragmentation process: The top panel shows the six Lagrangian neighbours of a given particle, the second panel illustrates how this particle accretes onto a neighbouring halo, the third panel depicts the merging of two haloes and successive accretion and, if there is no accretion, the particle is marked as belonging to a filament in the bottom panel.}\label{fig:cases}
\end{figure}

The hierarchical formation of objects is done due to the grouping of orbit-crossing particles into haloes by tracing the merging processes for each particle individually. Briefly, two main processes contribute to the hierarchical clustering: The first one is the accretion of particles onto haloes and the second one the merging of haloes. For this purpose, the particles of the realization are sorted in chronological order of collapse. Starting with the earliest collapse time the particles are assigned either to a halo or to filaments at the corresponding collapse times. In order for a collapsing particle to be accreted by a halo, it must fulfill a number of conditions. One of them is that the candidate halo must already contain one of its 6 nearest neighbours in the Lagrangian space of initial conditions. The fragmentation process contains 4 different cases which are sufficient for performing the identification and merging history of haloes (Fig. ~\ref{fig:cases}).

If none of the 6 Lagrangian neighbours have collapsed, then the particle is a local maximum of the inverse collapse time. This particle is a seed for a new halo having the unit mass of the particle and is created at the particle's position. Obviously, the earliest particle to collapse is the first halo.\\
In the case that the collapsing particle touches only one halo, then the accretion condition, if the halo is close enough, is checked. When the accretion condition is satisfied, then the particle is added to the halo, otherwise it is marked as belonging to a filament. Thus, not every collapsed particle becomes a relaxed halo but can be tagged as a filament. The particles that only touch filaments are marked as filaments as well. If the collapsing particle has more than one touching halo as Lagrangian neighbour, then the merging condition is checked for all halo pairs. Pairs that satisfy the conditions are merged. The accretion condition for the particle is checked for all touching haloes both before and after merging. In the case that the particle can accrete to both haloes, without the haloes merging, it accretes onto that halo for which the distance $d$ in units of halo size $R_N$ is smaller. It may happen that particles fail to accrete even though the haloes merge. \\If the collapsing particle does not accrete onto the candidate haloes in the case they are too far, it becomes a filament. But later for this filament particle there is still the possibility to accrete when its neighbour particle accretes onto a halo. This is done in order to mimic the accretion of filaments onto the haloes. Notice that up to 5 filament particles can flow into a halo at each accretion event.

\begin{figure*}
 \includegraphics[width=0.9\textwidth]{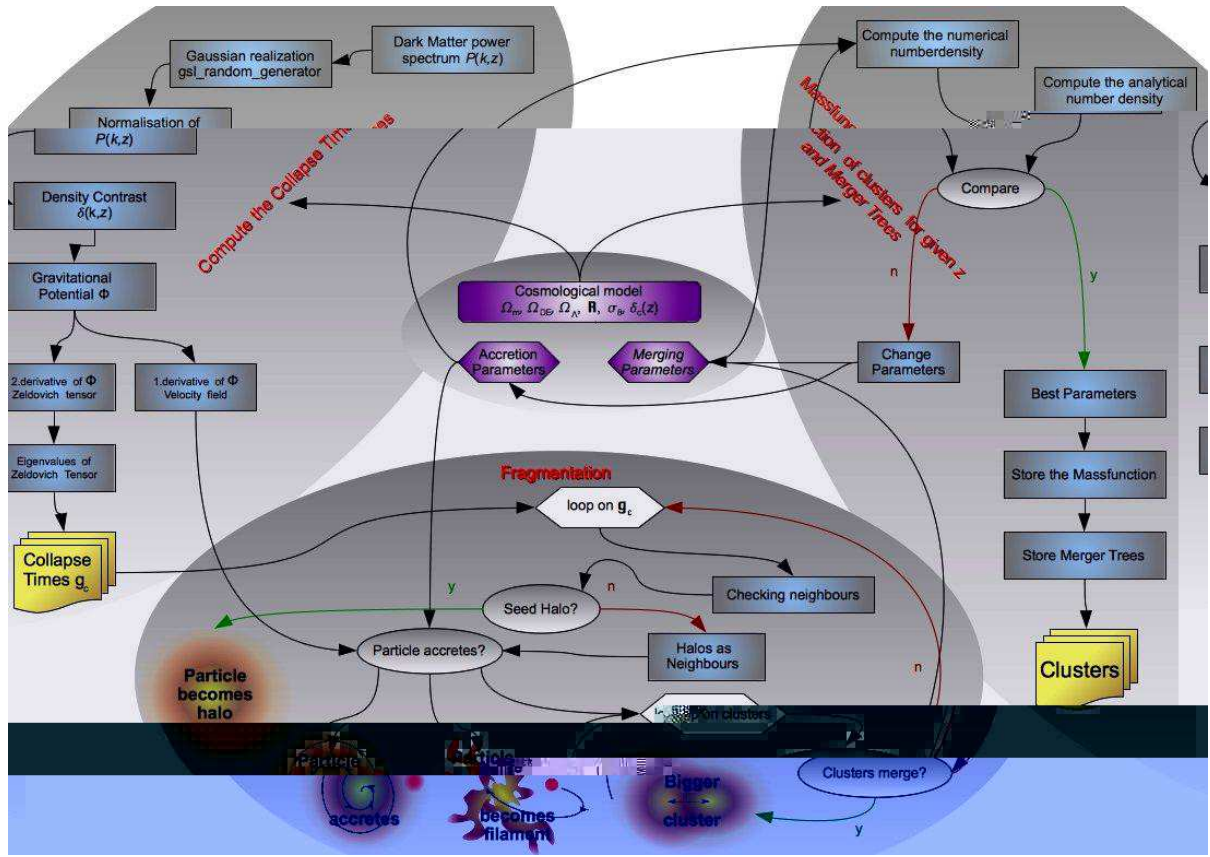}
  \caption{Flow chart of the code: The main 3 blocks of the code are shown. For a given set of parameters (centre) we compute the collapse times (top left) and for each collapsing particle we apply the fragmentation procedure (bottom). Finally we analyse the statistical properties of the fragmented objects (top right).}
\end{figure*}


\subsection{Velocity statistics}\label{sect_velstat}
In the Eulerian specification of the flow field, the flow quantities are depicted as functions of fixed position $\bmath{q}$ and time $t$. 
The peculiar velocity expressed in terms of the displacement field $\bmath{D}$ in the Lagrangian specification of the flow field would then be
\begin{equation}
\bmath{\upsilon}(t) =\dot{\bmath{x}}(t)= \dot g(t)\bmath{D}(\bmath{q})
\end{equation}
Now let us study the time evolution of two elements positioned at the Lagrangian coordinates $\bmath{q}_a$ and $\bmath{q}_b$. The Eulerian positions and peculiar velocities at these two points are then given as
\begin{eqnarray}
\bmath{x}_a\equiv\bmath{x}_a(\bmath{q}_a,t)=\bmath{q}_a+g(t)\bmath{D}(\bmath{q}_a)\equiv \bmath{q}_a + g(t)\bmath{D}_a,\\
\bmath{x}_b\equiv\bmath{x}_b(\bmath{q}_b,t)=\bmath{q}_b+g(t)\bmath{D}(\bmath{q}_b)\equiv \bmath{q}_b + g(t)\bmath{D}_b,\\
\bmath{\upsilon}_a(t)\equiv \bmath{\upsilon}_a(\bmath{q}_a,t)=\dot g(t)\bmath{D}_a,\\
\bmath{\upsilon}_b(t)\equiv \bmath{\upsilon}_b(\bmath{q}_b,t)=\dot g(t)\bmath{D}_b,
\end{eqnarray}
The relative velocity of the two particles is given by
\begin{equation}
\bmath{\upsilon}_{ab}=\bmath{\upsilon}_b(t)-\bmath{\upsilon}_a(t)=\dot g(t)(\bmath{D}_b-\bmath{D}_a)\equiv \bmath{\upsilon}_{\parallel}(t)+\bmath{\upsilon}_{\perp}(t),
\end{equation}
where $\bmath{\upsilon}_{\parallel}(t)$ and $\bmath{\upsilon}_{\perp}(t)$ stands for the components parallel and perpendicular to $\bmath{r}_{ab}(t)\equiv\bmath{x}_b(t)-\bmath{x}_a(t)$.
We can now compute the probability distribution function (PDF) of the pairwise peculiar velocity $\bmath{\upsilon}$ with separation $\bmath{s}$ from the initial PDF as \citep{1998ApJ...492..421S}:
\begin{eqnarray}
P(\bmath{\upsilon},\bmath{s},t) = 
\frac{1}{4\pi s^2}\int 4\pi r^2 dr\: 
d\upsilon_{\parallel i} d\upsilon_{\perp xi} d\upsilon_{\perp yi}\;\;p(\upsilon_{\parallel i},\upsilon_{\perp xi},\upsilon_{\perp yi};r)\nonumber \\
\times \delta(s-r_{ab}(t))\delta(\upsilon-\upsilon_{\parallel i}(t))
\end{eqnarray}
where $p(\upsilon_{\parallel i},\upsilon_{\perp xi},\upsilon_{\perp yi};r)$ is the initial PDF which depends only on $\bmath{\upsilon}_{\parallel i}$ and $\bmath{\upsilon}_{\perp i}\equiv\sqrt{\upsilon^2_{\perp xi}+\upsilon^2_{\perp yi}}$ where $\upsilon_{\perp xi}$ and $\upsilon_{\perp yi}$ are the two components of $\bmath{\upsilon}_{\perp i}$ perpendicular to each other. The subscript $i$ denotes quantities at some initial time.

\begin{figure}
\begin{center}
 \includegraphics[width=0.20\textwidth]{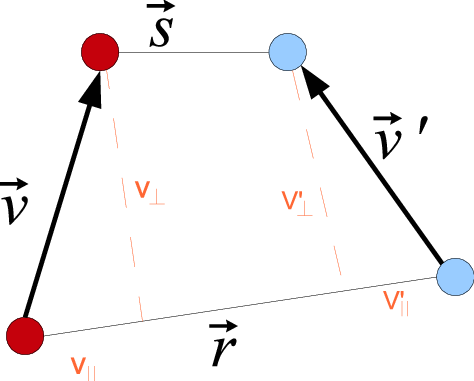}
  \caption{the visualisation of the configuration}
\end{center}
\end{figure}

Putting the Eulerian positions and peculiar velocities into the above PDF gives:
\begin{equation}
P(\bmath{\upsilon},\bmath{s},t)\propto\int^{\infty}_{r^{\star}}rdr \;\;p(\upsilon^{\star}_{\parallel i},\upsilon^{\star}_{\perp i};r),
\end{equation}
where $\upsilon^{\star}_{\parallel i}$ and $\upsilon^{\star}_{\perp i}$ stand for
\begin{eqnarray}
\upsilon^{\star}_{\parallel i} & \equiv & \frac{\dot g_i}{g}\left(\frac{s^2}{r}-\frac{sg}{r\dot g}\upsilon-r\right),\\
\upsilon^{\star}_{\perp i} & \equiv & \frac{\dot g_i}{g}s\left(1-\frac{s^2}{r^2}-\frac{g^2}{r^2\dot g^2}\upsilon^2+\frac{sg}{r^2\dot g}\upsilon\right),\\
r^{\star} & \equiv & \left|s-\frac{g}{\dot g}\upsilon\right|.
\end{eqnarray}
We have to specify the initial PDF $p(\upsilon^{\star}_{\parallel i},\upsilon^{\star}_{\perp i};r)$ in order to be able to compute the desired PDF $P(\bmath{\upsilon}, \bmath{s},t)$. As we were dealing with the longitudinal mode, the peculiar velocity in the linear regime is related to the $\delta(\bmath{x},t)$ and accordingly to its Fourier transform $\delta_{\bmath{k}}(t)$
\begin{equation}
\bmath{\upsilon}(\bmath{x},t)=i\frac{\dot g_i}{g_i}\int \frac{\bmath{k}}{k^2}\delta_{\bmath{k}}(t_i)e^{i\bmath{kx}}\frac{d^3k}{(2\pi)^3}.
\end{equation}
It is important to emphasize that the initial pairwise peculiar velocities are Gaussian distributed like the initial density fluctuations. From the velocity correlation tensor $\left<\upsilon_{a}\upsilon_{b}\right>$ one obtains the projections:
\begin{eqnarray}
\left<\upsilon_{\parallel }\upsilon_{\parallel }\right>&=&\sum_{a,b}\left(r^{\parallel}_ar^{\parallel}_b-\frac{1}{3}\delta_{ab}\right)\left<\upsilon_{a}\upsilon_{b}\right>,\\
\left<\upsilon_{\perp }\upsilon_{\perp }\right>&=&\sum_{a,b}\left(r^{\perp}_ar^{\perp}_b-\frac{1}{3}\delta_{ab}\right)\left<\upsilon_{a}\upsilon_{b}\right>.
\end{eqnarray}
Thus, the two-point correlation functions are given by \citep{1988ApJ...332L...7G}:
\begin{eqnarray}
\left<\upsilon_{\parallel i}\upsilon_{\parallel i}\right>&=&\frac{1}{6\pi^2}\left(\frac{\dot g_i}{g_i}\right)^2\int dkP_i(k)\left(1-3j_0(kr)+6\frac{j_1(kr)}{kr}\right)\\
\left<\upsilon_{\perp i}\upsilon_{\perp i}\right>&=&\frac{1}{6\pi^2}\left(\frac{\dot g_i}{g_i}\right)^2\int dkP_i(k)\left(1-3\frac{j_1(kr)}{kr}\right)
\end{eqnarray}
Finally, the initial probability distribution function is given by
\begin{equation}
p(\upsilon^{\star}_{\parallel i},\upsilon^{\star}_{\perp i};r)=\frac{e^{-T}}{\sqrt{(2\pi)^3}Y_{\parallel}(r)Y^2_{\perp}(r)},\;\;\;\;\;\;T\equiv\frac{(\upsilon^{\star}_{\parallel i})^2}{2Y_{\parallel}(r)}+\frac{(\upsilon^{\star}_{\perp i})^2}{2Y_{\perp}(r)}
\end{equation}
where $\left<\upsilon_{\parallel i}\upsilon_{\parallel i}\right>\equiv Y_{\parallel}(r)$ and $\left<\upsilon_{\perp i}\upsilon_{\perp i}\right>\equiv Y_{\perp}(r)$. Therefore, we can now obtain the desired PDF $P(\bmath{\upsilon},\bmath{s},t)$ through the integration $P(\bmath{\upsilon},\bmath{s},t)\propto\int^{\infty}_{r^{\star}}rdr \;\;p(\upsilon^{\star}_{\parallel i},\upsilon^{\star}_{\perp i};r)$. We relate the distance $\bmath{r}$ in the analytical model to the distance of two haloes at the time of merging. Furthermore we substitute for the power spectrum a filtered spectrum smoothed at a wavelength which corresponds to the halo mass.
Computing the velocity distribution at a distance $\bmath{s}$ corresponding to the merging condition yields distributions very similar to those obtained by PINOCCHIO. In particular we confirm the trend of steeper distributions at lower mass ratio.\\
As expected the velocity distribution for the analytical PDF is shallower than the distribution from PINOCCHIO data. This reflects the fact that merging processes in PINOCCHIO conserve momentum but not energy since merging is an inelastic collision. At each merging the velocity of the final halo is given by $\upsilon_f=(\upsilon_1m_1+\upsilon_2m_2)/(m_1+m_2)$ since the momentum is conserved. Because of the energy loss high velocities do not appear in the PINOCCHIO velocity distribution and therefore the curve decreases faster. But the loss of energy does not occur in the analytical configuration and therefore the curve is shallower.

\begin{figure}
\begin{center}
 \includegraphics[width=0.45\textwidth]{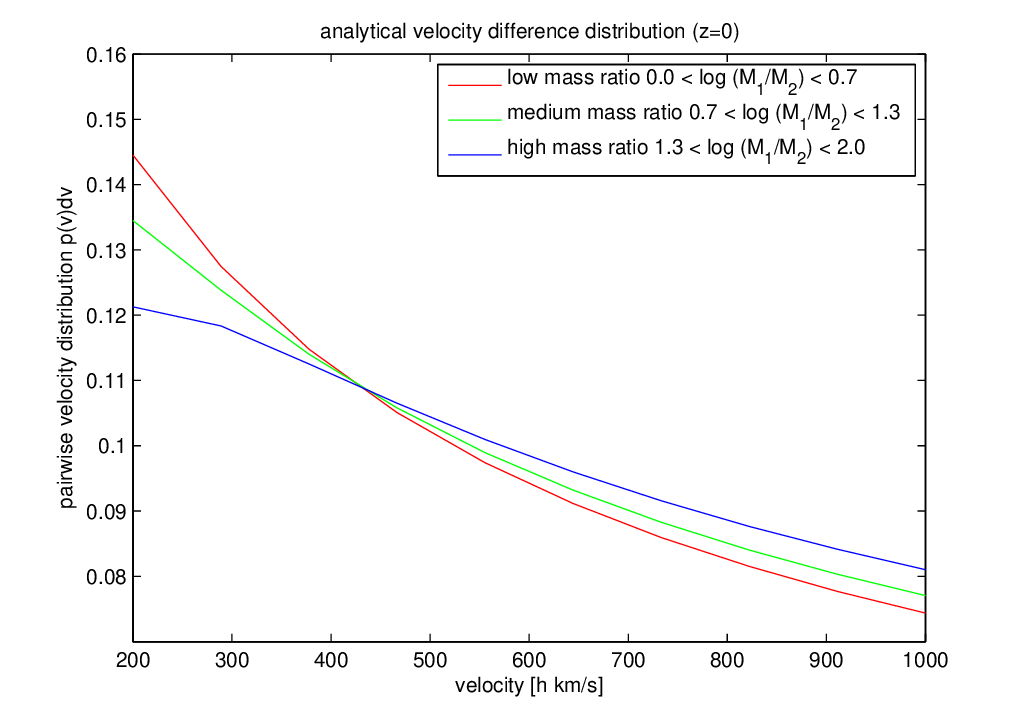}
  \caption{The analytical pair velocity probability density for three intervals in mass ratio: low mass ratio $0.0<\log (M_1/M_2)<0.7$ (red line), medium mass ratio $0.7<\log (M_1/M_2)<1.3$ (green line) and high mass ratio $1.3<\log (M_1/M_2)<2.0$ (blue line).}
\end{center}
\end{figure}

\section{results}
In the following we present the results from a simulation for $128^3$ particles in a box of side length $256~\mathrm{Mpc}/h$, carried out in the framework of Lagrangian perturbation theory with our C++ reimplementation of the PINOCCHIO code for following the merging activity of the large-scale structure. There is always the problem of the coherence of the velocity field on large scales. For that reason, large simulation boxes are always preferrable for the investigation of velocity statistics. Due to nonlinear mode-coupling even large modes influence the dynamics on small scales \citep{1997MNRAS.286...38C}, although this coupling decreases proportional to the inverse wavenumber or faster in classical perturbation theory. This difficulty can in principle be treated with PINOCCHIO using the method described by \citep{2005A&A...440..799M} which consists in adding large-scale modes to the Gaussian initial conditions, furthermore a volume of size $256~\mathrm{Mpc}/h$ has been demonstrated to sample the velocity field sufficiently well, as velocity differences are much less dependent on large-scale modes than velocities: Heuristically, the velocity difference statistic is closely related to that of the velocity field divergence, and has therefore the same properties as the density field itself, which is well sampled by our volume.

\begin{figure*}
\begin{center}
\begin{tabular}{cc}
\includegraphics[width=0.49\textwidth]{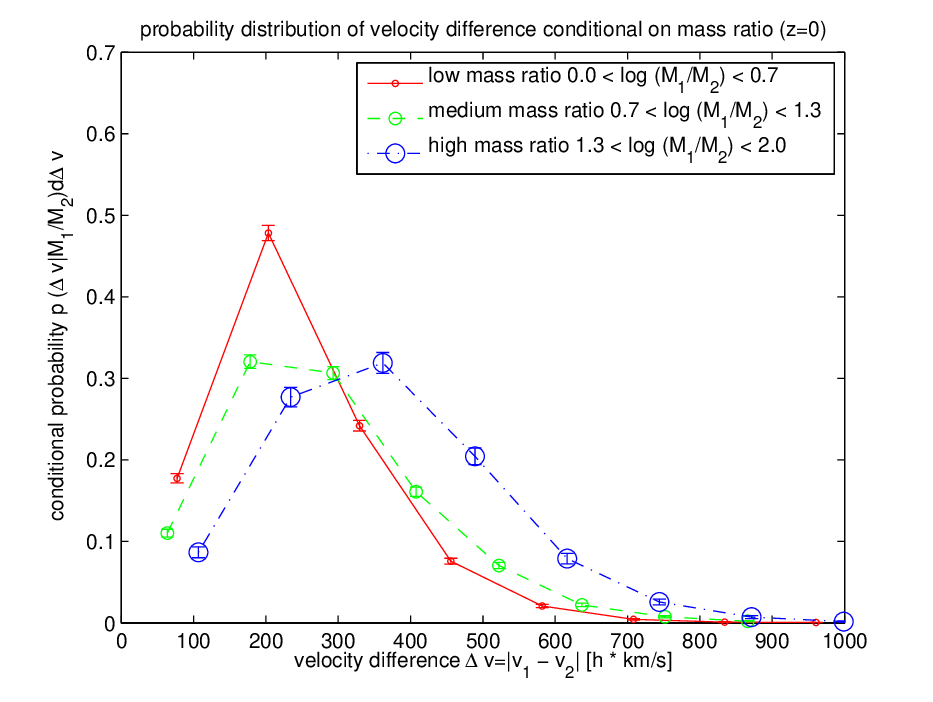} & \includegraphics[width=0.49\textwidth]{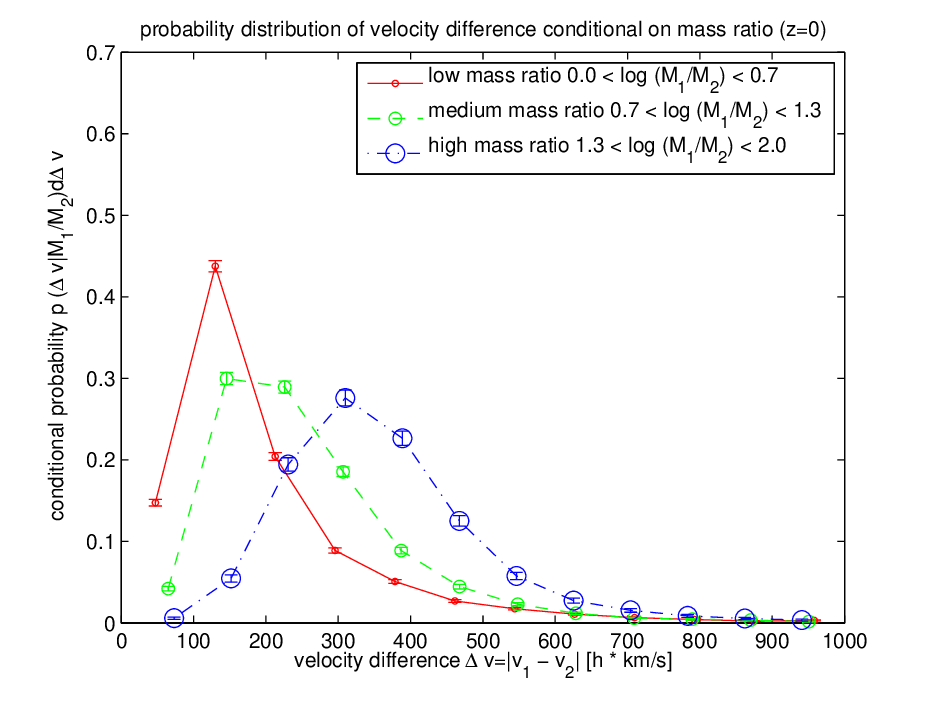} \\
\includegraphics[width=0.49\textwidth]{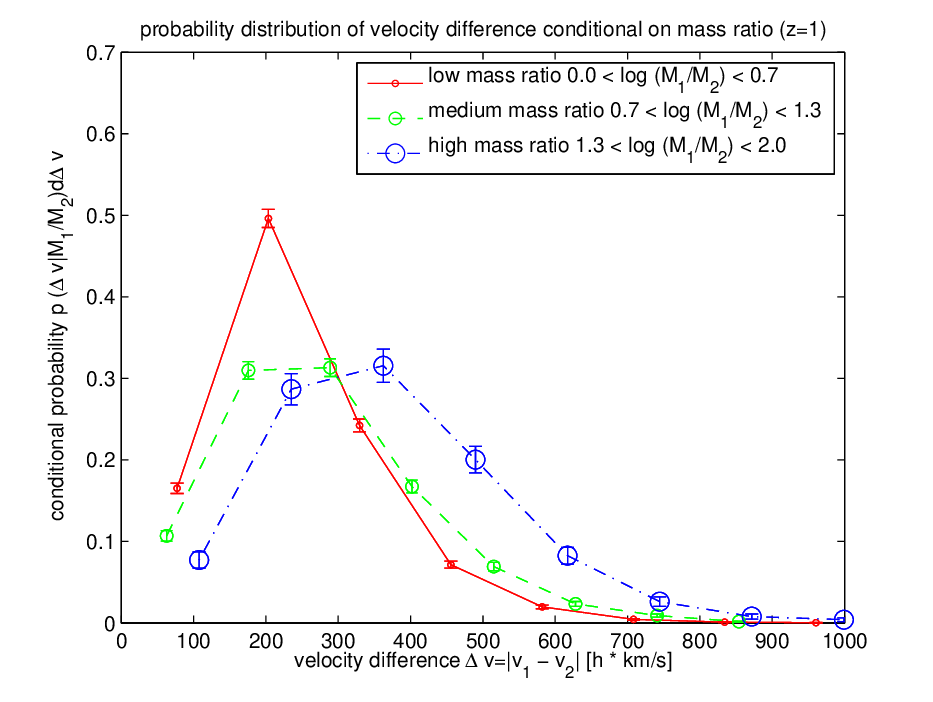} & \includegraphics[width=0.49\textwidth]{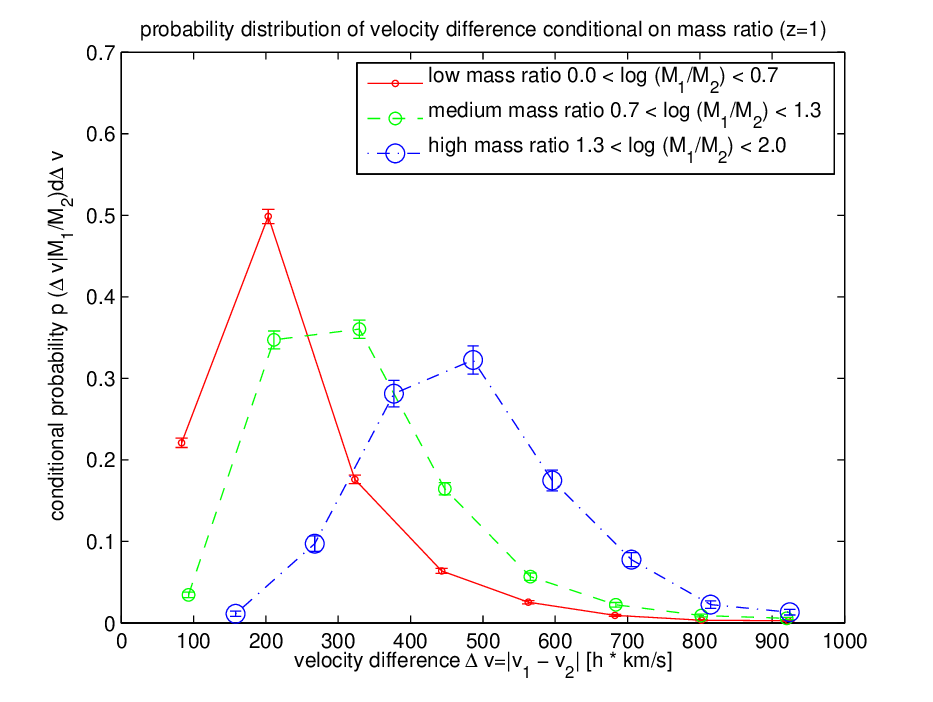} \\
\includegraphics[width=0.49\textwidth]{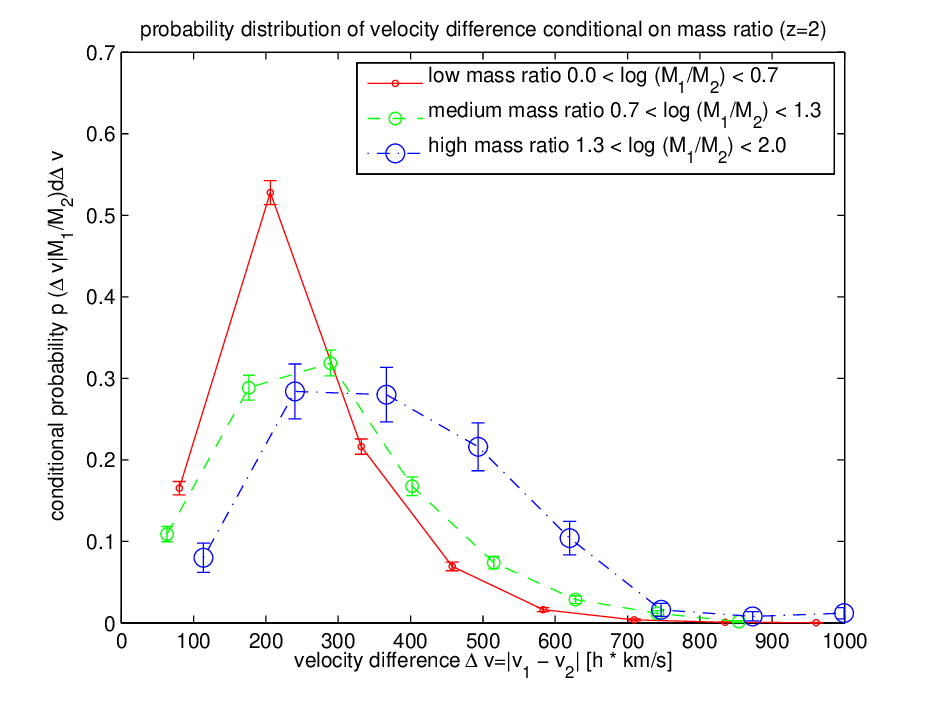} & \includegraphics[width=0.49\textwidth]{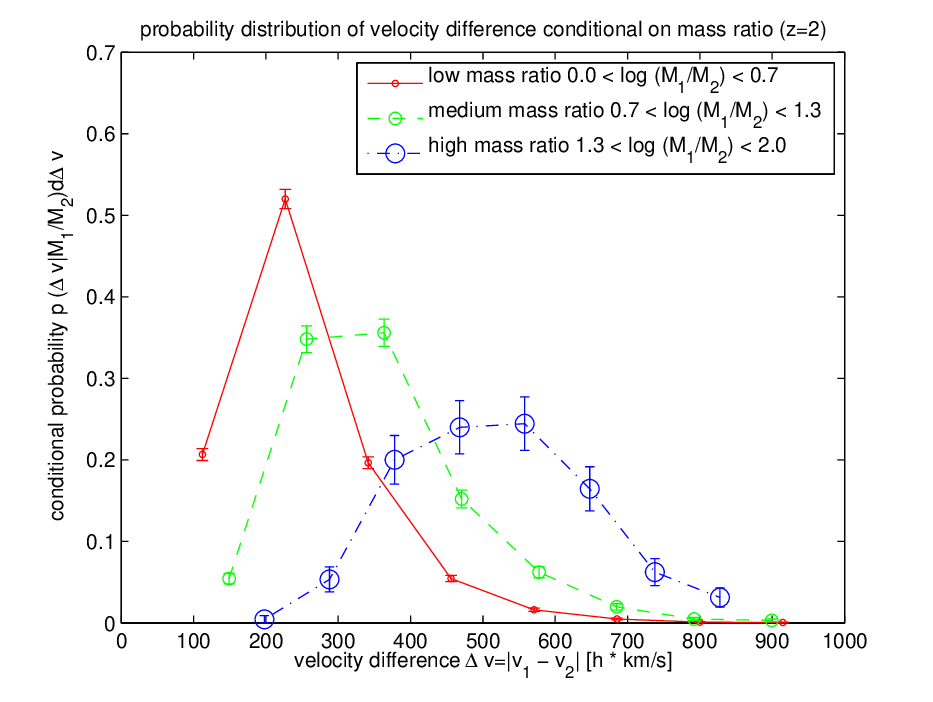}
\end{tabular}
\end{center}
\caption{Probability distributions of velocities conditional on halo mass ratio for redshift $z=0$ (first row), $z=1$ (second row) and $z=2$ (third row), comparing PINOCCHIO (left column) with the Millennium simulation (right column).}
\label{fig:PINmil}
\end{figure*}

The first step was to create a realisation of a Gaussian density field for a specified $\Lambda$CDM spectrum. From the density contrast the gravitational potential and Zel'dovich tensor were derived. Differentiations were performed again in Fourier space allowing us to recover the quantities with minimum noise. For each point $\bld{q}$ of the Lagrangian initial coordinates and each smoothing radius $R$, the collapse time, determined by the time at which the particle is predicted to enter a high-density multi-stream region, and ellipsoidal truncation were then computed using Lagrangian perturbation theory. Determining the eigenvalues of the Zel'dovich tensor, it is straightforward to calculate the collapse times evaluating the expressions of the individual orders~(\ref{eq:collapseTimes}). For each particle only the earliest collapse time is recorded with the corresponding smoothing radius and velocity. The third order Lagrangian prescription improves the collapse times. The first axis to collapse is the one corresponding to the smallest $\lambda_3$ eigenvalue indicating the convergence and the fact that the Zel'dovich approximation makes the largest contribution to collapse dynamics. The collapsed medium is then fragmented into isolated objects using the algorithm which we described above to mimic the accretion and merger events of hierarchical collapse. We distinguish between collapsed particles belonging to relaxed haloes or to lower-density filaments using the accretion and merging conditions. Once the fragmentation process is completed we studied the relative velocities of the haloes at merging.

\begin{figure*}
\includegraphics[width=0.85\textwidth]{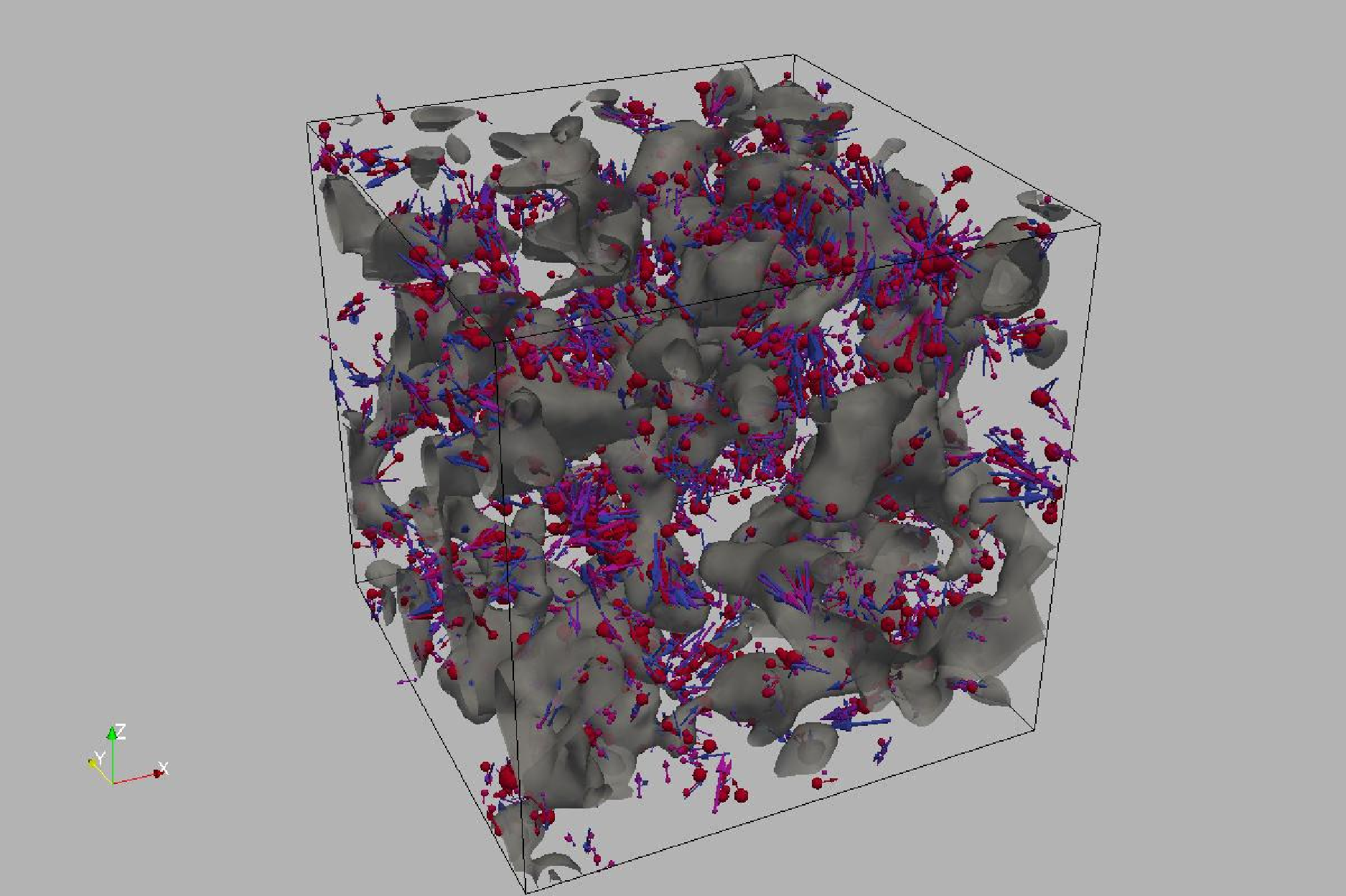}
\caption{Density field, smoothed with a Gaussian kernel of $8~\mathrm{Mpc}/h$, with contours at $-2\sigma$ and $-3\sigma$, superimposed on the velocity field visualised by arrows and the halo distribution, where the size of the spheres is indicative of the logarithmic halo mass.}\label{fig:pinocchio4}
\end{figure*}

\begin{figure*}
\includegraphics[width=0.85\textwidth]{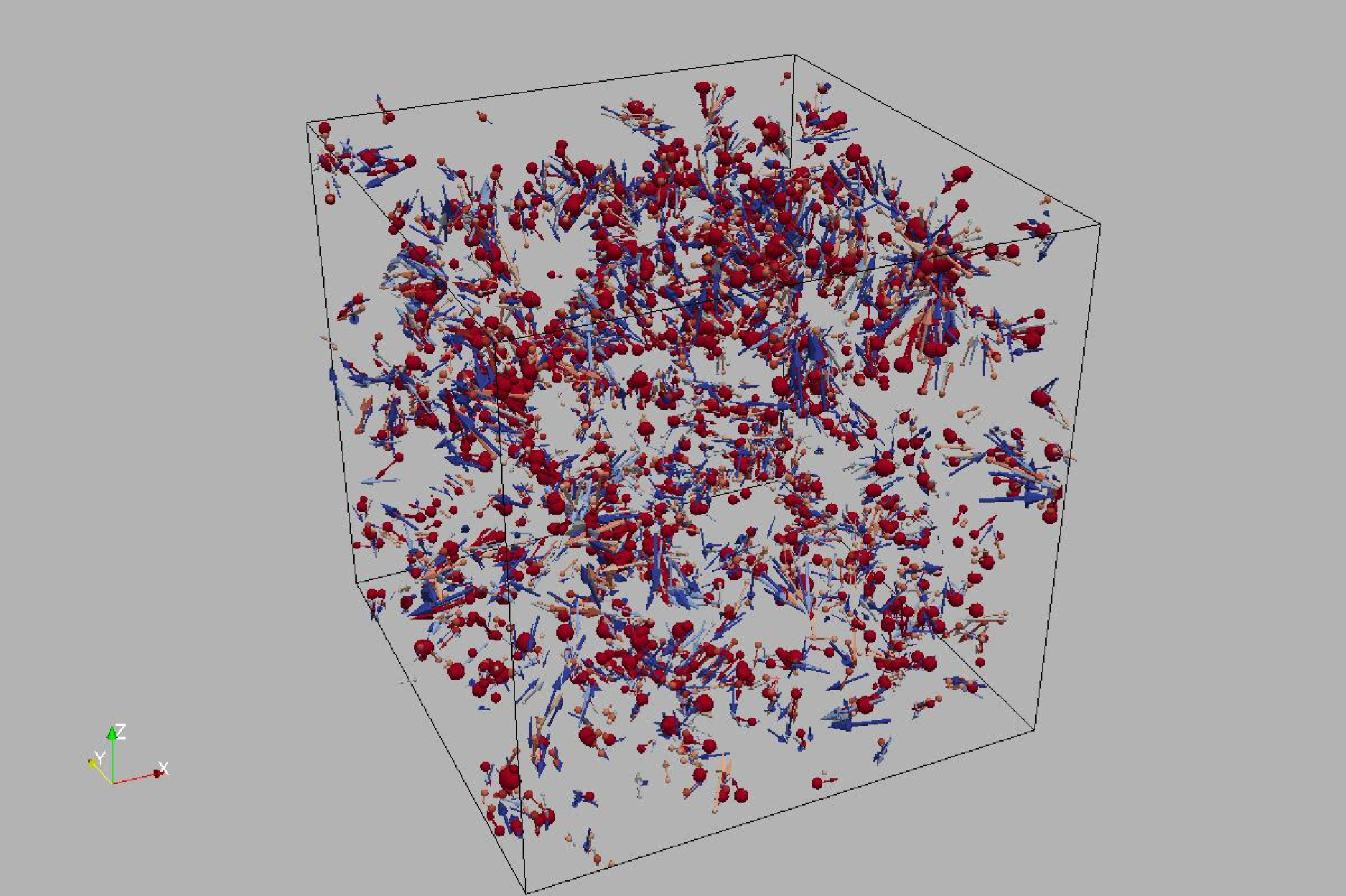}
\caption{Similar to Fig.~\ref{fig:pinocchio4} only without isodensity contours for making the velocity field more visible. Logarithmic halo mass is indicated by size, and the arrow length is proportional to the halo velocity.}\label{fig:velocity}
\end{figure*}

Figs.~\ref{fig:pinocchio4} and ~\ref{fig:velocity} illustrate structure formation dynamics in the Lagrangian picture: Matter is transported out of initially underdense regions and is accumulated in superstructures, where merging is prominent due to the high particle density. In fact, the most massive objects are found in regions of converging velocities, a nice example of which can be found in the upper right front corner of the simulation cube. One sees coherent flow patterns on the scale of the correlation length of the density field. In the vicinity of massive structures one can observe larger relative velocities compared to underdense regions. This can be traced back to the fact that the velocity field of an overdense region has a larger variance compared to the cosmological average. The statistics of the velocity field is translated to that of the haloes by imposing momentum conservation in the merging process. Therefore, the figure confirms the expectation of high pairwise velocities in overdense regions.

In order to compare our results with the Millennium Simulation \citep{2005Natur.435..629S} we have applied the same merging conditions to the Millennium data. In Figure~\ref{fig:PINmil} we compare the probability distribution of the relative velocity at the time of merging, for three different mass ratio intervals: approximately equal masses $\log(M_1/M_2)<0.7$, intermediate values for the mass ratio $0.7<\log(M_1/M_2)<1.3$ and high mass ratios $\log(M_1/M_2)>1.3$. The data is further split into three redshifts: $z=0,1,2$, while at higher redshifts the numbers of massive haloes is not sufficient for deriving the probability density as statistical error bars are too large to draw conclusions. The subdivision of the velocity range between 0-1000 km/s/h allows us to investigate the features of the velocity distribution while obtaining reasonable statistical error bars, and we normalise all histograms to unity. The error bars of the PINOCCHIO runs ($V=(256~\mathrm{Mpc/h})^3$) are small enough to allow quantitative conclusions and are comparable to those from the Millennium-data ($V=(256~\mathrm{Mpc/h})^3$). For drawing velocity distributions we selected halos from mass intervals of identical width ranging from $4\times10^{12}M_{\odot}/h$ to $1\times10^{14}M_{\odot}/h$. 

For controlling the time-stepping, GADGET (which was used to carry out the Millennium simulations) has a feature which synchronizes the individual time-steps for each particle prior to writing a simulation output at a pre-specified redshift. The Pinocchio simulation was stopped at the identical redshift, and given this approach, we do not expect difficulties to arise due to the time-discreteness of the Millennium output. Furthermore, as long as the structures are in the limit, where perturbation theory is applicable, which means that the trajectories of the particles can be extrapolated from the initial conditions, the choice of the time-stepping should not matter much, in contrast to nonlinear structures with large gradients perpendicular to the particle trajectory.

\begin{figure}
\includegraphics[width=0.45\textwidth]{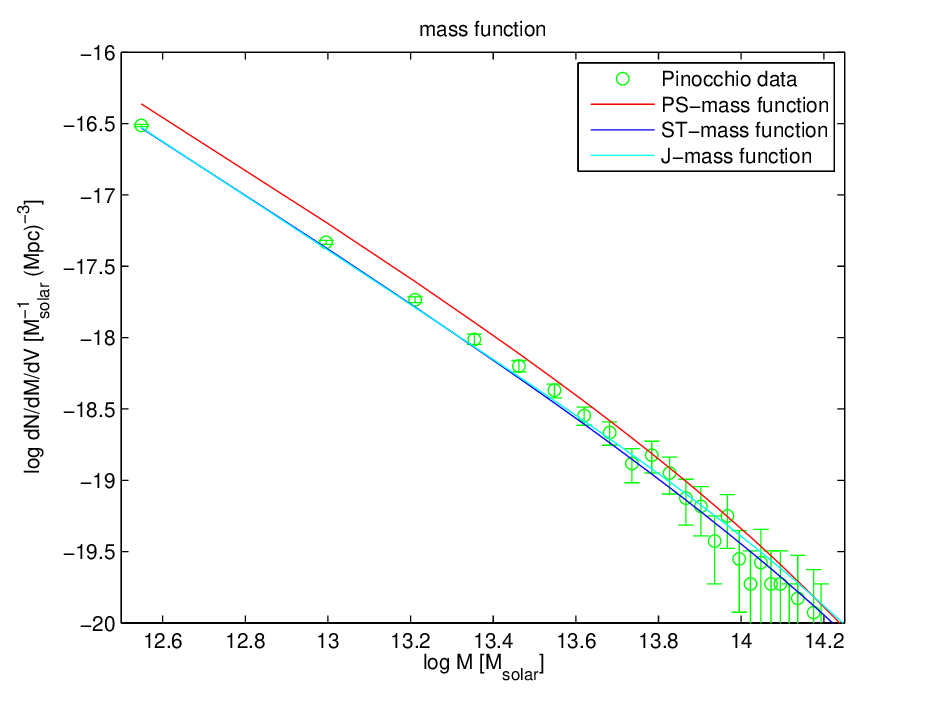}
\caption{comparison of the PINOCCHIO mass function with the analytical Press-Schechter, Sheth-Tormen and Jenkins mass functions} 
\label{fig:mf}
\end{figure}

A Metropolis--Hastings algorithm was used for checking the dependence of the mass function on the PINOCCHIO-parameters and we were able to reproduce mass functions in agreement with analytical functions for the proposed set of parameters. The influence of the parameter choice on the shape of the mass function was verified in our implementation, and the total number of haloes formed corresponds well to analytical expectations and the Millennium simulation.
For most of the tests presented in this paper we used a Opteron 285 2.4 GHz 64 bit computer with a total memory of 8GB. For example, resampling the initial conditions onto a $256^3$ grid, the first step of initialising the Gaussian random field and computing the orbit-crossing requires about 30 minutes of CPU time and the second part of identifying the haloes takes only a few additional minutes. Fig.~\ref{fig:mf} illustrates the agreement of the PINOCCHIO mass function with the analytical expectation.

Quite generally, the figures reproduce the basic behaviour expected from analytical arguments as outlined in Sect.~\ref{sect_velstat}. The distributions are steeper at lower mass ratio in both the Millennium data set as well as in PINOCCHIO. The shape of the distributions between PINOCCHIO and the Millennium simulation is very similar, particularly well at low redshifts, with the curves for the intermediate and the high mass ratio peaking at almost identical values for the relative velocity. At high redshifts PINOCCHIO underestimates the velocities in medium and high mass ratio mergers by a small amount. This underestimate is a known feature of Lagrangian codes and is discussed in the appendix of \citet{2005A&A...440..799M}. All curves terminate at velocities of $\simeq1000\mathrm{km/s}$ underlining the sparcity of high-velocity mergers \citep{2006MNRAS.370L..38H}. Naturally we expect the distributions of the Millennium simulation to drop faster than corresponding distributions from PINOCCHIO, because the latter treats merging processes as an inelastic collision  and does not follow the dissipative dynamics inside haloes. Given the good agreement, we believe that halo velocity catalogues for a number of applications such as redshift space distortions, large-scale bulk flows and merging processes can be reliably derived from Lagrangian codes.

\section{Summary}\label{sect_summary}
The aim of this paper is an investigation of merging processes in the cosmic large-scale structure in the framework of Lagrangian perturbation theory and to compare the results for the pairwise velocity distribution obtained with an adaptation of the PINOCCHIO algorithm to $n$-body data.
\begin{itemize}
\item{It is comparatively simple in PINOCCHIO to construct merger trees as it is not necessary to identify haloes in the particle distribution and as one can directly follow the merging processes between haloes, such that the haloes in PINOCCHIO correspond to friends-of-friends particle groups in $n$-body simulations.}
\item{Halo properties, such as the distribution of masses, spins and now also merger trees and pairwise velocities can be reliably derived from a Lagrangian code, with the differences in the distribution being smaller than the statistical error bars. Hereby, we have investigated the derivation of the mass function from PINOCCHIO and used similiar parameters as quoted by \citet{2002MNRAS.331..587M} for our studies of the velocity distributions. We have also investigated that the velocities show the correct scaling with cosmological parameters, and the correct scaling with halo mass ratio in comparison to an analytical calculation.}
\item{In comparison to an analytical pairwise velocity distribution, PINOCCHIO is able to reproduce the trend of shallower distributions with increasing halo mass ratio. At high velocities, however, PINOCCHIO exhibits a steeper behaviour compared to that predicted by the analytical calculation at fixed mass ratio, which is explained by the fact that merging processes in PINOCCHIO are treated as inelastic collisions with conserved momentum but not conserved energy. Because of this energy loss, high velocities are not present in PINOCCHIO data and the distribution is steeper.}
\item{We find a general agreement between the velocity distributions of PINOCCHIO and the Millennium simulation, both in terms of relative numbers and values for the absolute velocity. Additionally, the scaling with redshift and mass ratio between merging haloes behaves very similarly. The peaks of the distributions at low redshifts coincide with each other, and although distributions from the Millennium simulation terminate earlier, this feature is not unexpected as PINOCCHIO does not treat the dissipative dynamics of merging haloes. Again, the correct dependence of pairwise velocity with halo mass ratio is recovered.}
\item{PINOCCHIO, relying on a phenomenological description of the merging process of two haloes and combining the individual momenta in an inelastic collision is very fast compared to $n$-body codes, which allows sweeps in the parameter space relevant to peculiar velocities, i.e. $\Omega_m$, $\sigma_8$ and the dark energy parameters, for which we have verified the basic relations expected by linear structure formation.}
\end{itemize}
Further questions include the environment-dependence of velocity statistics, which is comparatively easy to do in Lagrangian perturbation theory. A very useful discriminant for this purpose is the number of positive eigenvalues of the shear tensor. One would expect smaller velocities inside voids and larger velocities in supercluster regions. In fact, this dependence can already be seen in Fig.~\ref{fig:pinocchio4}. Other extensions include the investigation of two-point statistics of the velocity field, and to answer questions related to velocity statistics \citep{1995MNRAS.272..447R}. In a future paper, we will investigate the dependence of the strong lensing cross-section of merging systems with velocities drawn from PINOCCHIO simulated volumes, and its dependence on the choice of cosmological parameters.

Anaglyphic versions of Figs.~\ref{fig:pinocchio4} and~\ref{fig:velocity} are available on request from the authors.

\section*{Acknowledgements}
We thank Volker Springel for his suggestions and access to the Millennium data base. LH's work is partially funded by the SNF. BMS's work is supported by the Graduate School of Fundamental Physics in the framework of the DFG's Excellence Initiative. MB receives funding from DFG's Transregio TR33. We appreciate the comments from our reviewer, Pierluigi Monaco. We would like to thank Francesco Pace for answering questions on the original FORTRAN code and Jean-Claude Waizmann for his suggestions to improve the figures.

\bibliography{bibtex/aamnem,bibtex/references}
\bibliographystyle{mn2e}

\appendix

\bsp

\label{lastpage}

\end{document}